\documentclass[aps,prc,twocolumn,groupedaddress,showpacs]{revtex4}

\usepackage{graphicx}

\begin{document}

\title{High Density Neutron Star Equation of State from 4U 1636-53 Observations}

\author{Timothy S. Olson}
\email[]{tim_olson@skc.edu}
\affiliation{Salish Kootenai College, PO Box 117, Pablo, MT 59855}

\date{\today}

\begin{abstract}
A bound on the compactness of the neutron star in the low mass x-ray binary 4U 1636-53 is used to estimate the equation of state of neutron star matter at high density. Observations of 580 Hz oscillations during the rising phase of x-ray bursts from this system appear to be due to two antipodal hot spots on the surface of an accreting neutron star rotating at 290 Hz, implying the compactness of the neutron star is less than 0.163 at the 90\% confidence level. The equation of state of high density neutron star matter estimated from this compactness limit is significantly stiffer than extrapolations to high density of equations of state determined by fits of experimental nucleon-nucleon scattering data and properties of light nuclei to two- and three-body interaction potentials.

\end{abstract}


\maketitle

\section{introduction}
Oscillations in x-ray brightness during bursts have been observed from ten low mass x-ray binaries with the \textit{Rossi} X-ray Timing Explorer \cite{Strohmayer2001}. These oscillations appear to be due to rotational modulation of one or a pair of antipodal hotspots on the surface of the accreting neutron star \cite{Strohmayer-etal-1997}. With this model the amplitude of the oscillations is determined by the mass-to-radius ratio of the neutron star, the compactness $M/R$. A more compact neutron star results in greater deflection of the x-ray photons emanating from a hot spot, making the hot spot visible to an observer for a larger portion of the rotation period and so reducing the amplitude of the brightness oscillations. The bounds on the compactness of the neutron star are particularly limiting when two antipodal hot spots are present \cite{MillerAndLamb1998}.

The source 4U 1636-53 appears to be the best candidate for providing a stringent limit on the compactness of neutron stars from x-ray burst oscillations \cite{Nath-etal-2001}. A 290 Hz subharmonic of the stronger 580 Hz brightness oscillation has been observed in five bursts from this source \cite{Miller1999}, but the subharmonic has not been confirmed in subsequent bursts \cite{Strohmayer2001}. The existence of the subharmonic would suggest the spin frequency of the neutron star is 290 Hz, and the presence of two antipodal hot spots is responsible for the 580 Hz modulation. A pair of high frequency quasiperiodic oscillations separated by 251 Hz has also been reported for this source and interpreted in a beat frequency model to be a consequence of a spin frequency of 290 Hz with two hot spots on the neutron star \cite{Miller-etal-1998}. The x-ray oscillations during the rising phase of bursts from 4U 1636-53 have been recently modeled as due to a pair of circular antipodal hot spots, each having an angular size that grows linearly in time \cite{Nath-etal-2001}. The two hot spot model constrains the neutron star compactness to $M/R < 0.163$ at 90\% confidence (using $G = c = 1$), requiring the radius of a 1.4 $M_\odot$ neutron star to be greater than 12.7 km. The 99\% confidence level compactness limit, $M/R < 0.183$, implies $R > 11.3$ km when $M = 1.4 M_\odot$.


During a burst event in low mass x-ray binaries the oscillation frequency is usually observed to evolve asymptotically to a limit that probably corresponds to the spin rate of the neutron star. Measurement of the asymptotic frequency at different orbital phases of the binary could reveal a Doppler modulation that would allow for a determination of the masses of the neutron star and the companion. A mass measurement together with the compactness limit would tightly constrain the radius of the star. A Doppler modulation has not been observed for 4U 1636-53 in an analysis of 26 burst oscillations \cite{Giles-etal-2001}. The absence of the modulation likely limits the neutron star mass in this system to $M < 1.6 M_\odot$ \cite{Giles-etal-2001}.

The purpose of this paper is to use the 90\% confidence level compactness bound on the 4U 1636-53 neutron star to estimate the high density equation of state of neutron star matter. Ideally one would have both a measured mass and radius for a number of neutron stars over a wide range of masses up to the maximum mass value. The observed mass-radius relationship could then be used to uniquely calculate the equation of state by inverting the Oppenhemier-Volkoff equation, the relativistic stellar structure solution to Einstein's equations \cite{Lindblom1992}. Even if the mass and radius are known for only a single neutron star, the high density equation of state can still be accurately estimated up to the central density yielding the mass of this star. At low density the equation of state can be determined by fitting experimental nucleon-nucleon scattering data and properties of light nuclei to two-body and three-body interaction potentials \cite{Wiringa-etal-1988, Akmal-etal-1998}. Confidence in these low density equations of state is high near normal nuclear density (energy density = 153 MeV/fm$^3$, baryon density = 0.16 fm$^{-3}$) because nuclear physics experiments allow for verification in the low density regime. Estimates of the maximum possible mass of neutron stars based on these equations of state have regarded them as experimentally verified up to twice normal nuclear density \cite{KalogeraAndBaym1996, Olson2001}. Assuming one has values of both the mass and radius for a single neutron star, the equation of state in the density range between the experimentally verified low density regime and the high central density of this star can be accurately estimated to have a polytrope form provided the matter does not undergo a phase transition in this density range \cite{Lindblom1992}. The polytrope high density equation of state is chosen to match the energy density and pressure of the low density equation of state at the upper limit of applicability of the low density equation, and the adiabatic index of the polytrope is chosen to yield a stellar radius corresponding to the observed radius of the star when the mass of the model star equals the observed mass. The resulting polytrope is an estimate of the equation of state of neutron star matter in the high density regime.

In the case of the neutron star in 4U 1636-53 we do not have measurements of both its mass and radius, but only an upper bound on the compactness and an upper limit on the mass. Nevertheless the high density equation of state is still tightly constrained by the compactness limit provided it is assumed the neutron star mass is at least $1.4 M_\odot$, a plausible assumption for an old accreting system like 4U 1636-53. In the next section the high density equation of state is estimated from the 90\% confidence level compactness bound on the neutron star in 4U 1636-53 assuming the neutron star mass is at least $1.4 M_\odot$. The main result of this paper is that the A18 + $\delta v$ + UIX* equation of state of Ref. \cite{Akmal-etal-1998} (hereafter the APR equation of state), a recently developed equation of state representative of the class of equations of state derived from fits of experimental nuclear physics data to interaction potentials, is too soft at high density to yield a star with the observed low compactness of the neutron star in 4U 1636-53. Further if the APR equation of state is used up to twice normal nuclear energy density, and any polytrope equation of state having a sound speed less than the speed of light is used above twice normal nuclear density, the model star compactness still exceeds the compactness of the 4U 1636-53 neutron star. Using the APR equation of state up to just a value in the range 1.55-1.64 times normal nuclear energy density and a stiff polytrope with an adiabatic index $\Gamma$ of 5.5 $\alt \Gamma \alt$ 7 at high density can produce a star with the low compactness of 4U 1636-53 without the speed of sound waves in the neutron star interior exceeding the speed of light.

\section{estimation of the high density equation of state}
The equation of state of high density neutron star matter is estimated in this section from the compactness bound on the neutron star in 4U 1636-53. The equation of state at low density is assumed to be known for energy densities less than a transition density $\tilde{\rho}$. Here the APR equation of state is used for the low density portions of the stellar interior, from normal nuclear density $\rho_{nn}$ up to $\tilde{\rho}$. The BPS equation of state \cite{Baym-etal-1971} is used for densities below $\rho_{nn}$ to model the matter outward through the neutron star crust to the surface. Recent work on determining the maximum possible mass of neutron stars has assumed the APR equation is accurate up to twice normal nuclear energy density \cite{KalogeraAndBaym1996, Olson2001}. For densities greater than $\tilde{\rho}$ an equation of state of polytrope form is used in this paper:
\begin{equation} \label{polytrope}
\frac{p}{\tilde{p}} = \left( \frac{\rho}{\tilde{\rho}} \right)^\Gamma \quad .
\end{equation}
The polytrope form can be expected to be a good approximation to the actual high density equation of state of the 4U 1636-53 neutron star provided the matter has not undergone a phase change to a softer condensed phase, such as a kaon condensate or a deconfined quark phase, at some density above $\tilde{\rho}$ \cite{Lindblom1992}. The compactness limit on the 4U 1636-53 neutron star requires the stellar radius to be large, so it is unlikely this star contains a condensed phase.

The algorithim for determining the high density equation of state begins by choosing the transition energy density $\tilde{\rho}$ where one switches between the high and low density equations of state. The pressure $\tilde{p}$ at the transition density is computed from the low density equation of state and the transition energy density value: $\tilde{p} = p \left( \tilde{\rho} \right)$. The adiabatic index $\Gamma$ in the polytrope high density equation of state Eq. (\ref{polytrope}) is treated here as a free parameter that is varied from small values (soft equations of state) to progressively larger values (stiffer equations of state) to find the set of model stars satisfying the 4U 1636-53 neutron star compactness bound.

The interior structure of each model star is assumed to be described accurately by the Oppenheimer-Volkoff solution, the solution to Einstein's equations for the interior of a nonrotating star in general relativity:
\begin{equation} \label{Op-V} 
\frac{dp}{dr}=-\frac{\left( \rho + p \right)\left[m(r)+4\pi r^3 p \right]}{r\left[ r-2m(r) \right]} \quad ,
\end{equation}
where
\begin{equation} \label{stellarMass}
 m(r) = \int_0^r 4\pi \rho \left( \xi \right) \xi^2 d\xi \quad .
\end{equation}
Rotational corrections to the Oppenheimer-Volkoff solution are small except for very rapidly rotating stars \cite{Friedman-etal-1986} and will be ignored here. Families of model stars, one family for each $(\tilde{\rho},\Gamma)$ pair, are determined by integrating the Oppenheimer-Volkoff solution Eq. (\ref{Op-V}) for a set of central energy density values. The polytrope equation of state Eq. (\ref{polytrope}) is used for densities exceeding $\tilde{\rho}$. The APR equation of state is used for densities below $\tilde{\rho}$ and above normal nuclear density. The BPS equation of state is used below $\rho_{nn}$.

\begin{figure}
\includegraphics{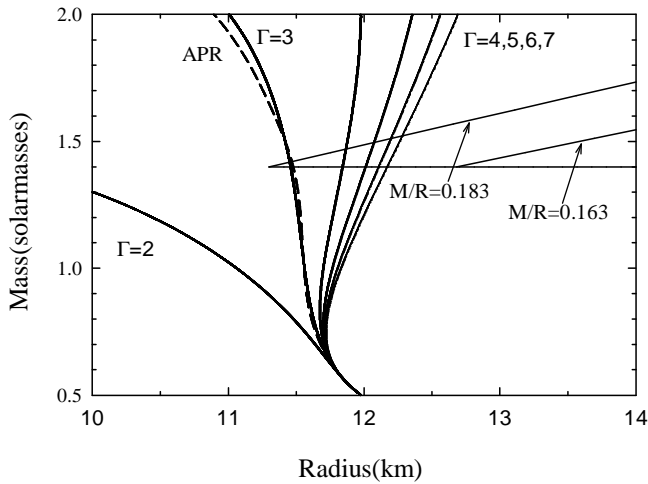}
\caption{\label{Fig_1} Neutron star mass vs. radius for the APR equation of state used up to twice normal nuclear energy density. The solid curves are for a polytrope high density equation of state, and the dashed curve is for the APR equation of state used at all densities. None of these models satisfy the 90\% confidence level 4U 1636-53 neutron star compactness limit, $M/R < 0.163$.}
\end{figure}

Figure \ref{Fig_1} shows how the neutron star mass depends on the radius for the case where $\tilde{\rho} = 2\rho_{nn}$. Several families of model stars are represented differing by the value of the adiabatic index of the high density polytrope. Also shown is the neutron star mass-radius curve resulting when the APR equation of state is used at all densities. Lines representing the 90\% and 99\% confidence level 4U 1636-53 neutron star compactness bounds delineate the allowed regions of the mass-radius parameter space. The allowed regions are below these compactness limit lines and above the 1.4$M_\odot$ line. The compactness of all these model stars exceeds the 90\% confidence level 4U 1636-53 neutron star compactness bound, $M/R < 0.163$, even for very stiff polytropes. Models with $\Gamma > 7$ are not physically realistic because the equation of state predicts the speed of sound waves,
\begin{equation}
v_s = \sqrt{\frac{\Gamma p}{\rho}} \quad ,
\end{equation}
exceeds the speed of light near the stellar center. Hence model stars with the APR equation of state used up to twice normal nuclear density and a polytrope at high density are ruled out at the 90\% confidence level. High density polytropes that are at least moderately stiff ($\Gamma \agt 3$) are consistent with the 99\% confidence level bound. The model using the APR equation of state at all densities is also consistent with the 99\% confidence level bound, $M/R < 0.183$, but is ruled out at 90\% confidence.

\begin{figure}
\includegraphics{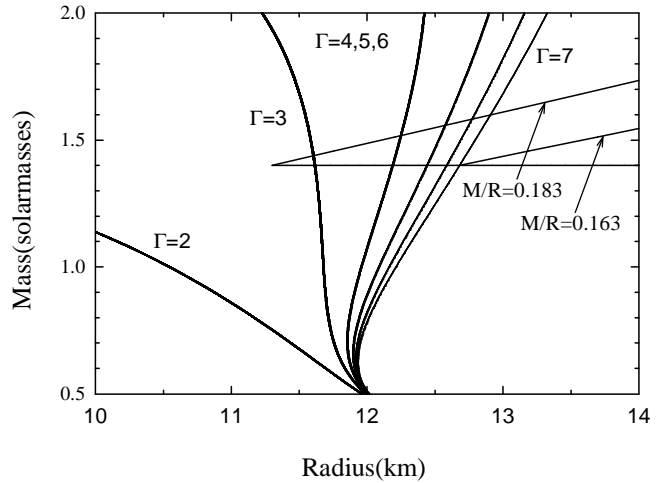}
\caption{\label{Fig_2} Neutron star mass vs. radius for the APR equation of state used up to 1.64 times normal nuclear energy density and a polytrope equation of state used for higher densities. Models with $\Gamma \approx 7$ satisfy the 90\% confidence level 4U 1636-53 neutron star compactness limit, $M/R < 0.163$, without the sound speed exceeding light speed.}
\end{figure}

Model stars satisfying the 90\% confidence level compactness bound can result if the transition energy density is chosen to be $1.64\rho_{nn}$ or less. For $\tilde{\rho} > 1.64\rho_{nn}$ the speed of sound at the stellar center exceeds the speed of light for all model stars having an adiabatic index value large enough to yield a compactness below the 90\% confidence level bound, and therefore are ruled out. For $\tilde{\rho} \leq 1.64\rho_{nn}$ the 90\% confidence level bound can be met without the sound speed exceeding the speed of light. Figure 2 shows neutron star mass vs. radius curves for the case where $\tilde{\rho} = 1.64\rho_{nn}$. The radius is near 12.7 km for a 1.4 $M_\odot$ star for the model families having $\Gamma$ near 7, and the sound speed is subluminal throughout the star provided $\Gamma < 7.023$. For $\Gamma = 7.023$ the sound speed just reaches the speed of light at the center of the star. The central energy density and central pressure of the 1.4 $M_\odot$, 12.7 km star are 317 MeV/fm$^3$ and 45.1 MeV/fm$^3$, respectively, when $\tilde{\rho} = 1.64\rho_{nn}$ and $\Gamma = 7.023$.

\begin{table}
\caption{\label{EOS} Equation of state of neutron star matter using the APR equation of state up to 1.64$\rho_{nn}$ and a $\Gamma = 7.023$ polytrope above 1.64$\rho_{nn}$.}
\begin{ruledtabular}
\begin{tabular}{cc}
energy density $\left( \frac{\text{MeV}}{\text{fm}^3} \right)$ & pressure $\left( \frac{\text{MeV}}{\text{fm}^3} \right)$\\
\hline
153 & 2.38\\
160 & 2.76\\
180 & 3.94\\
200 & 5.17\\
220 & 6.43\\
240 & 7.86\\
260 & 11.27\\
280 & 18.97\\
300 & 30.80\\
320 & 48.46\\
\end{tabular}
\end{ruledtabular}
\end{table}

\begin{figure}
\includegraphics{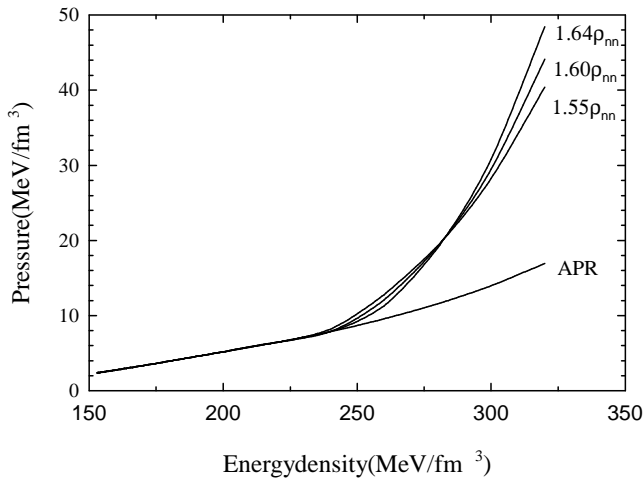}
\caption{\label{Fig_3} Comparison of four equations of state, three that use the APR equation of state at low density and a high density polytrope (labeled with the value of the transition energy density), and the fourth that uses the APR equation of state at all densities (labeled APR). The adiabatic index of each polytrope is chosen to yield a 1.4 $M_\odot$, 12.7 km neutron star: $\Gamma = 7.023$ for $\tilde{\rho} = 1.64\rho_{nn}$, $\Gamma = 6.25$ for $\tilde{\rho} = 1.60\rho_{nn}$, and $\Gamma = 5.55$ for $\tilde{\rho} = 1.55\rho_{nn}$. The equations of state utilizing a high density polytrope differ little from each other but are significantly stiffer than the APR equation of state at high density.}
\end{figure}

Table \ref{EOS} lists particular values of the pressure as a function of the energy density for the $\tilde{\rho} = 1.64\rho_{nn}$, $\Gamma = 7.023$ equation of state. Model stars that have $M = 1.4 M_\odot$, $R$ = 12.7 km, and subluminal sound speeds are also obtained for $\tilde{\rho} < 1.64\rho_{nn}$. For $\tilde{\rho} = 1.60\rho_{nn}$, $\Gamma = 6.25$ in order to make the 1.4 $M_\odot$ star have a radius of 12.7 km. The 1.4 $M_\odot$, 12.7 km star has $\Gamma = 5.55$ when $\tilde{\rho} = 1.55\rho_{nn}$. Figure \ref{Fig_3} compares the equations of state for four cases: $\tilde{\rho} = 1.64\rho_{nn}$ with $\Gamma = 7.023$, $\tilde{\rho} = 1.60\rho_{nn}$ with $\Gamma = 6.25$, $\tilde{\rho} = 1.55\rho_{nn}$ with $\Gamma = 5.55$, and the APR equation of state used at all densities. The three equations of state that use a high density polytrope differ very little from each other. Hence the equation of state of Table \ref{EOS} can be regarded as an estimate of the minimally stiff equation of state of neutron star matter satisfying the 4U 1636-53 90\% confidence level neutron star compactness bound that also maintains a subluminal sound speed throughout the stellar interior. A stiffer equation of state is implied if the 4U 1636-53 neutron star is actually more massive than 1.4 $M_\odot$. Figure \ref{Fig_3} also shows the equations of state utilizing a high density polytrope are significantly stiffer than the APR equation of state at high density.

\section{discussion}
A new equation of state (Table \ref{EOS}) valid for describing neutron star matter up to the central density of a 1.4 $M_\odot$ star has been developed in this paper based on observations of the low mass x-ray binary 4U 1636-53. The validity of this equation of state rests on three main assumptions: the actual compactness of the 4U 1636-53 neutron star does not exceed the 90\% confidence level bound, the mass of the 4U 1636-53 neutron star is near 1.4 $M_\odot$, and the APR equation of state is valid up to near 1.64 times normal nuclear energy density. The APR equation of state used at all densities is sufficient to describe the structure of the 4U 1636-53 neutron star if its compactness is actually near the 99\% confidence level bound, $M/R = 0.183$, rather than near the 90\% confidence level bound value of $M/R = 0.163$. The APR equation of state used at all densities is also adequate to meet the 90\% confidence level bound if the 4U 1636-53 neutron star is 1.27$M_\odot$ or less. (The APR equation of state is seen from Fig. \ref{Fig_1} to yield a radius of 11.5 km for a mass of 1.27$M_\odot$, making the compactness of this star $M/R = 0.163$.) If the mass of the 4U 1636-53 neutron star is larger than 1.4$M_\odot$, the equation of state of neutron star matter must be even stiffer than the Table \ref{EOS} equation of state. The APR equation of state is based on a large body of experimental evidence at low density, so it is reasonable to trust its validity up to $\tilde{\rho} \approx 1.6\rho_{nn}$. Figure \ref{Fig_3} shows the equation of state derived here (Table \ref{EOS}) is relatively insensitive to the exact transition energy density value as long as $\tilde{\rho}$ is near 1.6$\rho_{nn}$. 

If the APR equation of state is used up to $1.55\rho_{nn} \alt \tilde{\rho} \alt 1.64\rho_{nn}$, a stiff polytrope, having an adiabatic index of 5.5 $\alt \Gamma \alt$ 7, is needed to produce a model star with a compactness $M/R = 0.163$ without the speed of sound waves in the neutron star interior exceeding the speed of light. The adiabatic index would be small, near 5/3 for nonrelativistic nucleons and 4/3 in the relativistic limit, if the pressure is due mainly to the Fermi kinetic energy. The adiabatic index is near 2 if the equation of state is dominated by static two-body interactions, near 3 if static three-body interactions are most important, and in excess of 3 if repulsive momentum-dependent interactions are dominant \cite{Akmal-etal-1998}. The large adiabatic index of the equation of state at high density implied by the 4U 1636-53 neutron star 90\% confidence level compactness bound suggests the APR equation of state underestimates at high density the strength of the repulsive momentum-dependent interactions.

\begin{acknowledgments}
This work was supported by the U.S. National Aeronautics and Space Administration through Grant No. NAG5-9541.
\end{acknowledgments}

\bibliography{lmxb}

\end{document}